\documentclass[%
 aip,
  pof, % Physics of Fluids substyle of AIP, no titles in citations
 amsmath,amssymb,
  ,%
]{revtex4-1}
\usepackage{CJK}

\usepackage{ulem}
\usepackage[T1]{fontenc}
\usepackage{calligra}

\usepackage{graphicx}% Include figure files
\usepackage{dcolumn}% Align table columns on decimal point
\usepackage{bm}% bold math

\usepackage{times}
\usepackage{verbatim}
\usepackage{graphicx}
\usepackage{graphics,epsfig}

\usepackage{theorem}
\usepackage{makeidx}
\usepackage{amsmath}
\usepackage{epic}
\usepackage{amscd}
\usepackage[dvips]{color}
\usepackage{siunitx}
\usepackage{lipsum}
\usepackage{mathrsfs}
\usepackage{bm}

\usepackage{bbm}
\usepackage{xy}
\usepackage{amssymb}
\usepackage{xcolor}
\usepackage{cancel}
\begin{document}

\title{%On local structures of (non-)ideal flows of two-fluid plasmas and neutral fluids
%{\color{red}Invariant geometric objects of Navier-Stokes and two-fluid plasma dynamics}
Local invariants in non-ideal flows of neutral fluids and two-fluid plasmas}
\author{Jian-Zhou Zhu}
\email{jz@sccfis.org}
\affiliation{
Su-Cheng Centre for Fundamental and Interdisciplinary Sciences, Gaochun, Nanjing, China
}%

%
%\author{B. Author}
% \homepage{http://www.Second.institution.edu/~Charlie.Author.}
%\affiliation{%
%Second institution and/or address%\\This line break forced% with \\
%}%

%\author{C. Author}%
% \email{}
%%  \altaffiliation[Also at ]
%% {}%Lines break automatically or can be forced with \\
%\affiliation{
%%\\This line break forced with \textbackslash\textbackslash
%}%

\date{\today}% It is always \today, today,

\begin{abstract}
The main objective is the locally invariant geometric object of any (non-ideal) fluid, while more attentions are paid to the untouched dynamical properties of two-fluid fashion. Specifically, local structures, beyond the well-known `frozen-in' to the barotropic flows of the generalized vorticities, of the two-fluid model of plasma flows are presented. More general non-barotropic situations are also considered. A modified Euler equation [T. Tao, Ann. PDE \textbf{2}, 9 (2016)] is also accordingly analyzed and remarked from the angle of view of two-fluid model, with emphasis on the local structures. And, the local constraints of high-order differential forms such as helicity, among others, find simple formulation for possible practices in modelling the dynamics. Thus, the Cauchy invariants equation [N. Besse and U. Frisch, J. Fluid Mech. \textbf{825}, 412 (2017)] may be enabled to find applications in non-ideal flows. Some formal examples are offered to demonstrate the calculations, and particularly interestingly the two-dimensional-three-component (2D3C) or the 2D passive scalar problem presents that a locally invariant $\Theta = 2\theta \zeta$, with $\theta$ and $\zeta$ being respectively the scalar value of the `vertical velocity' (or the passive scalar) and the `vertical vorticity', may be used as if it were the spatial density of the globally invariant helicity, providing a Lagrangian prescription to control the latter in some situations of studying its physical effects in rapidly rotating flows (ubiquitous in atmosphere of astrophysical objects) with marked 2D3C vortical modes or in purely 2D passive scalars.
\end{abstract}

%\pacs{47.27.Gs, 52.30.Gz, 51.10.+y}

\maketitle

\section{Introduction}
Constructing material invariants moving with the fluid, especially from known ones of ideal flows, such as the local helicity characterizing the spiral (magneto)hydrodynamic [(M)HD] spatial structure, has been of particular interests \cite{KuzminPLA83} (see also %Oseledets
Ref. \onlinecite{OseledetsCotMMS89}, and, in any odd dimensions, %Gama and Frisch
Ref. \onlinecite{GamaFrisch93}). No non-trivial practical application was presented, but the original ideas have become the important source of theoretical developments. With the application of differential forms \cite{Flanders89}, the problem can be unified with the notion of `Lie/\textit{local} invariant'.  Unlike the helicity characterizing global topology \cite{Moffatt69}, spatial integral of locally invariant 3-form helicity is of course invariant regardless the boundary conditions, for the moving material domain. In general, invariant local helicity is not the spatial density of the invariant global helicity $\mathscr{H}$ in a fixed domain, but, as we will see in a two-dimensional-three-component (2D3C) or 2D passive scalar problem, a locally invariant quantity may work as if it were the spatial density of $\mathscr{H}$.

Recently, the `finite-time blowup' issue of the Euler equation has gotten some illumination from models preserving some of the original local and global invariance properties \cite{TaoAnnPDE16}; and, rewriting the local invariance laws of the conservative systems in Lagrangian coordinates under appropriate conditions\cite{BesseFrischJFM17} appears helpful for various issues of fundamental fluid mechanics. We observe that the so-called Cauchy invariants equation of the latter \cite{BesseFrischJFM17} actually has nothing to do with the mechanisms leading to the local invariance laws. And, we know that some of the ideas and properties (such as conservation and symmetry laws) of ideal flows are useful for studying non-ideal ones, calling for theories and techniques for constructing invariants of the latter. For example, relevant to the global invariant, the Gaussian method results in tractability in the statistics of some model dynamics \cite{ChernovEyinkLebowitzSinaiJMP93}. The Gauss-Navier-Stokes for `equivalent' turbulence ensemble \cite{GallavottiBook14}, for instance, with specified global invariants however are in general in the Eulerian framework, lacking the scenario in the generalized Lagrangian (i.e., Lie) framework with local invariants: parallel efforts should be beneficial. So, it deserves to develop ideas and techniques relevant to Lie invariants for non-ideal flows.

Since one of our motivations for this study was to go beyond the `weak excitation approximation' and ideal treatment for aero-acoustic energy partition affected by helicity \cite{ZhuJFM16}, and beyond our previous ideal extended MHD calculations of solar wind chirality \cite{ZhuMNRAS17} to prepare to address the fine chiral structures of compressible, instead of incompressible \cite{hydrochirality}, two-fluid plasmas (e.g., Tur and Yanovsky \cite{TurYanovskyJFM93}; see also Sagdeev et al. \cite{SagdeevMoiseevTurYanovskii86} in the context of strong turbulence and topological soliton), it is natural to turn to the non-ideal local dynamics (with multiplicative invariants constructed from the Lie-varying forms, as mentioned above). Indeed, although `frozen-in' to the \textit{barotropic} flows of the generalized vorticity in the two-fluid model of plasma flows is well-known, it is desirable to find more precise relations, such as those similar to the (extended) MHD ones established in Besse and Frisch \cite{BesseFrischJFM17}, of the generalized vorticity and helicity about the local structures for such special and other more general non-barotropic situations. And, with the Lie formulation, it may be useful to apply the local constraints of high-order forms such as helicity, among others, to model the dynamics, with possibly some kind of generalized (measure-valued, for instance) solutions; that is, using the `nice' properties of ideal classical solutions to constrain the `turbulent' solutions in some modeling or parametrization approaches.

\section{Local invariants and the generalised Cauchy invariants equations for two-fluid dynamics
}

For a neutral fluid with the $1$-form $V$ corresponding to the velocity vector $v$ [c.f. Eq. (\ref{eq:2fluid}) below in terms of vectors for the two-fluid model] in the Riemannian $n$-manifold ($n=3$ in our discussions, if not otherwise specified) endowed with the metric tensor $g_{\alpha \beta}$ ($=\delta_{\alpha \beta}$ in the Euclidean $\mathbb{R}^n$ beyond which this note does not really bother to go into the general curved manifolds, except for denoting the generality of the discussions in some situations) and with the volume form $d\verb"vol"=\sqrt{|det(g_{\alpha \beta})|}dx^1\wedge...\wedge dx^n$, we have
\begin{equation}\label{eq:neutral}
\partial_t V + L_v V = -dh + d(V,V)_g/2 \ \text{with barotropic entropy} \ dh = dp/\rho,
\end{equation}
as in Tao \cite{TaoAnnPDE16} (in Euclidean space) and in Besse and Frisch \cite{BesseFrischJFM17}, some of whose and of standard textbooks' notation conventions, such as the upper (respectively lower) cases for forms (respectively vector) and the inner product $(\bullet,\bullet)_g$ with respect to the metric $g$, have been followed.
It follows directly from Eq. (\ref{eq:neutral}) that the 2-form vorticity $\Omega=dV$ (inversely $V=d^{\star}\Delta^{-1} \Omega$) satisfies
\begin{equation}\label{eq:neutralOmega}
\partial_t \Omega + L_v \Omega = 0,
\end{equation}
for which terminologies such as `frozen-in', `Lie invariant' and `Lie advection/transport', among others, can be found in the literature (we will also somewhat arbitrarily use for the general form in the position of $\Omega$ `Lie-carry' or simply `invariant', among other variants, when no confusion would arise.)
Tao \cite{TaoAnnPDE16} replaces in the above the Lie derivative $L_v$ with $L_u$ with respect to the other fluid velocity vector $u$ and the corresponding 1-form velocity $U=d^{\star}A \Omega$ with $A$, a general `vector potential operator', replacing the Hodge Laplacian $\Delta$ (from the Biot-Savart law). Such a generalized Euler system has been shown to lose the `global (in time) regularity' (allow `no classical solution for all time') for some $A$ even such chosen as to preserve several conservation laws (see details in Ref. \onlinecite{TaoAnnPDE16} for the precise mathematical set up and meanings). Such models are of `two-fluid' fashion [for reference,
one can get the `dynamical' equation
\begin{eqnarray}\label{eq:TaoU}
\partial_t U + d^{\star}A L_u dV=0
\end{eqnarray}
for the complementary fluid by performing the time derivative on both sides of $U=d^{\star}A \Omega$ and by using Eq. (\ref{eq:neutralOmega})] and we will come back to this point later.

One can add a `gauge' $G$ to the equation for Weber's transformation function $w$,\cite{Weber1868}
\begin{equation}\label{eq:Weber}
(\partial_t + L_v)w = -h + (V,V)_g/2 + G,% of Besse and Frisch \cite{BesseFrischJFM17}
\end{equation}
%for the Weber transformation function $w$
to form the Lie-invariant helicity $(V-dw) \wedge dV$, with
\begin{equation}\label{eq:WeberGauge}
dG \wedge dV = 0, \ \text{e.g.,} \ \Omega=dV = dG \wedge D \ \text{for some $1$-form $D$}.
\end{equation}
That is, the results of Kuz'min \cite{KuzminPLA83} and Oseledets \cite{OseledetsCotMMS89} (without our $G$), beyond which is the generalised Cauchy invariants equation and formula of Ref. \onlinecite{BesseFrischJFM17} obtained, can be more general; actually, even more general in the sense that the Lie-source/sink --- the right hand side --- of the momentum equation (\ref{eq:neutral}) does not need to be closed (not to mention the exactness). And, the Weber transformation is not always necessary, as we will show in Sec. \ref{sec:LieVarying}.

%Relevant to fine-scale structure of plasma dynamics, Ref. \onlinecite{BesseFrischJFM17} has studied the extended magnetohydrodynamics and also pointed out that it does not reduce to the classical single-fluid magnetohydrodynamics when the ion and electron skin depths vanish. There may be also intrinsic limitations in the model itself with the simplifications in the philosophy of `MHD'. Two-fluid description of plasma dynamics without the (generalised) Ohm's-law simplification may still be desirable in more general situations.

%\subsection{Flows of two-fluid plasmas}

The two-fluid model of a plasma reads in the familiar lower-case/vector form
\begin{equation}\label{eq:2fluid}
m_s \rho_s\frac{dv_s}{dt} = -\nabla p_s + q_s \rho_s (e + v_s \times b),
\end{equation}
for charged ($q_s$) ion and electron species $s$ of mass $m_s$ and density $\rho_s$. To be more explicit for discussing the MHD idea and the (generalized) Ohm's law, we write down the equation for ion fluid including the mutual friction (but not the internal viscosity) with the electron fluid:\cite{Braginskii}
\begin{equation}\label{eq:ion}
m_i \rho_i\frac{dv_i}{dt} = -\nabla p_i + q_i \rho_i (e + v_i \times b) - m_i \rho_i \verb"v"/\tau,
\end{equation}
with $\verb"v"$ being the velocity difference between the ion and electron fluids, the electric current, and $\tau$ the mutual collision time between the positively and negatively charged particles. Then, the combined momentum equation (from those of the two fluids) and the induction equation, from modeling the electron fluid momentum equation [(generalized) Ohm's law] united with the Maxwell equations, constitute the (extended) MHD. The (generalized) Ohm's law approximation is not always controllable and it should be clarified whether and how the fundamental, such as the local/Lie-structure, of the dynamics is related to the original/full two-fluid model: Note that the latter model is sometimes also (imprecisely) called `two-fluid MHD' in the literatures, including one of ourselves \cite{hydrochirality}, but actually it is `electro-magneto-hydro-dynamics' with the electric field/force also explicitly present.

We first show that, after some manipulations, especially applying the Hodge decomposition twice, respectively for collecting the time and Lie derivatives of the magnetic potential $A$ in the generalised momentum $P_s = m_s V_s+q_s A$ with $dA = B$, Eqs. (\ref{eq:2fluid}) can be organized into a fashion common to those studied by Besse and Frisch \cite{BesseFrischJFM17} written in differential forms, besides Maxwell's equation \cite{Flanders89}:
\begin{equation}\label{eq:2fluidP}
  (\partial_t + L_{v_s}) P_s = -d[ h_s - (v_s,v_s)_g/2 - K_s ] + \mathfrak{h}_s, \ \text{with barotropic} \ dh_s = dp_s/\rho_s.% \ \text{(assuming barotropicity)},
\end{equation}
[Barotropicity will eventually be abandoned in later discussions.] We have in the above united the Gauss equation $dB = 0$ (absence of magnetic monopole) with Cartan's magic formula $L_v \Omega = i_v d\Omega + d(i_v \Omega)$ (for any form $\Omega$ acting on which the inner product $i_v$ and the exterior derivative entangle to result in the Lie derivative) and its derivative $dL_{v_s} = L_{v_s}d$. Thus, $L_{v_s} A - i_{v_s}B$ is closed,
\begin{equation}\label{eq:GaussCartan}
d (L_{v_s} A - i_{v_s}B) = 0,
\end{equation}
and we can transform the Lorentz force
\begin{equation}\label{eq:LorentzHodge}
\text{$-q_s i_{v_s}B$ (the dual $1$-form of the vector field $q_s v_s \times b$) into $-q_s L_{v_s} A$},
\end{equation}
up to the arbitrary $0$-form $K_{sL}$ and harmonic $1$-form $\mathfrak{h}_{sL}$ from the Hodge decomposition $L_{v_s} A - i_{v_s}B = dK_{sL} + \mathfrak{h}_{sL}$ (the third co-exact form disappears due to the closeness), the latter combined with the other one dealing with the electric force in upper-case/1-form (dual to the velctor $q_s e$):
\begin{equation}\label{eq:Faraday}
\text{Faraday's law} \ \partial_t B + dE = 0 \ \text{implies} \ d(\partial_t A + E) = 0 \ \text{and} \ \partial_t A + E = dK_t + \mathfrak{h}_t,
\end{equation}
transforming the electric force to the time derivative of the magnetic potential. So, in Eq. (\ref{eq:2fluidP}),
\begin{equation}\label{eq:HodgeDecomposition}
%\text{the arbitrary $
K_s = q_s(K_{t} + K_{sL}) \ \text{and} \ %$ and the harmonic $
\mathfrak{h}_s = q_s(\mathfrak{h}_{t} + \mathfrak{h}_{sL}).%$}.
\end{equation}

Then, assuming $\mathfrak{h}_s \ne 0$%_{\bullet} = 0$ (zero Betti number $b_1$, as is the case for simply connected manifolds)
, or, even for non-barotropic case, it is direct to introduce the two-fluid Weber transformation function $w_s$ by
\begin{equation}\label{eq:Pi}
(\partial_t + L_{v_s}) w_s = dp_s/\rho_s - \mathfrak{h}_s %h_s - d(v_s,v_s)_g/2 - K
\end{equation}
to form $\Pi_s = P_s + w_s$.  [The coupled dynamics of entropy $(\partial_t + L_{v_s})\eta_s = 0$ (adiabatic as the flow is `ideal') and mass\cancel{e} $(\partial_t + L_{v_s})M_s = 0$, however, are not explicitly involved in the derivation of the Lie-carried $2$-form generalised vorticity $\Omega_s$, similar to the situation of compressible neutral fluids.\cite{BesseFrischJFM17}] % which is Lie-carried by $v_s$.
And, for $d\Pi_s = \Omega_s$, we have
\begin{equation}\label{eq:2fluidO}
  \partial_t \Omega_s + L_{v_s} \Omega_s = 0.% \ \text{with} \ dP_s = \Omega_s.
\end{equation}
Note that the barotropic frozen-in property of the generalized vorticity corresponding to the Lie-invariance of $dP_s$ in such a special situation.
The `Cauchy invariants equation' follows with a pullback $\varphi_{st}^*$ (of the flow generated by $v_s$)
\begin{equation}\label{eq:2FluidCauchy}
d\Pi_{sk} \wedge dx_s^k = \varphi_{st}^*\Omega_s = \Omega_{s0}.
\end{equation}

Ref. \onlinecite{BesseFrischJFM17} has been able to recognise and generalise the classical Cauchy invariants equation to $(\partial_t + L_v)d\alpha = 0$ for any $(p-1)$-forms $\alpha \in \Lambda^{p-1}(\mathcal{R})$ with $\mathcal{R} \subset M$ being a bounded region of the manifold $M$%(Riemannian structure supplemented with a metric $g$ is not necessary if the Hodge dual is not used to obtain the Cauchy formula)
:
\begin{equation}
\label{eqn:GCI}
\frac{1}{(p-1)!}\,\delta_{j_1\ldots j_{p-1}}^{i_1\ldots i_{p-1}} d\alpha_{i_1\ldots i_{p-1}}\,
\wedge dx^{j_1}\wedge \ldots \wedge dx^{j_{p-1}}=\varphi_t^* d\alpha = d\alpha_0,
\end{equation}
with $x = \varphi_t$ and $\dot{\varphi}_t = v$, and, $\delta_{j_1\ldots j_{p-1}}^{i_1\ldots i_{p-1}}$ being the generalised Kronecker.

To construct higher-order invariant (in particular, the local helicity/spirality) and to apply the above result, the simplest and conventional way is to start with two easily-found Lie invariants and to construct the `multiplicative' one from the wedge product of the known ones (assuming exactness of the product, otherwise special care would be needed). The invariant local helicity of an invariant $1$-form results trivially from the fact that the exterior derivative of a invariant is still invariant, and so is their wedge product. The original momentum is in general not Lie invariant, thus some kind of `gauge' is introduced. For the local helicity to have a close relation with the momentum $P_s$, one hopes that at least the exterior derivative of the latter is invariant, which is the case when the Lie-source/sink of $P_s$ is closed. When $\mathfrak{h}_{\bullet} = 0$ (zero Betti number $b_1$, for contractibility of the domain, say) or $d\mathfrak{h}_s = 0$, and, when there exists $h_s$ satisfying $dh_s=dp_s/\rho_s$, the Lie-source/sink of $P_s$ is exact, or at least closed, thus the invariance of $dP_s$. If the Lie-source/sink for $P_s$ is exact, one can gauge $P_s$ by an exact form $w_s = d\mathscr{G}_s
$ with precisely the opposite Lie-source/sink: For instance, following Eq. (\ref{eq:2fluidP}), $\Pi_s = P_s + w_s$ with closed $G_s$ in
\begin{equation}\label{eq:exactGauge}
(\partial_t + L_{v_s}) \mathscr{G}_s = h_s - (v_s,v_s)_g/2 - K_s + G_s,
\end{equation}
instead of Eq. (\ref{eq:Pi}), won't change the invariant $2$-form vorticity, i.e., the exterior derivative $dP_s = d\Pi_s$ (otherwise not), while $\Pi_s$ being also invariant. Then, the 3-form local self-helicities $\sigma_s = \Pi_s \wedge \Omega_s$ (and their Hodge duals $\star \sigma_s$) are invariant.\cite{SteinhouerIshida98PoP} Working in 3-manifolds with $d \sigma_s =0$ and assuming zero Betti number $b_3 = 0$, we have exactness $\sigma_s = d \xi_s$, and thus, the Cauchy invariants equation
\begin{equation}\label{eqn:Cauchy2fluid}
  \frac{1}{2} \delta_{ij}^{kl} d \xi_{skl} \wedge dx_s^{i} \wedge dx_s^{j} = \sigma_{s0}.
\end{equation}

After formulating the two-fluid model in differential forms with Eq. (\ref{eq:2fluidP}), or for the more general adiabatic case as gauged by Eq. (\ref{eq:Pi}), we have in the above applied the ideas and techniques of Ref. \onlinecite{BesseFrischJFM17} for those simpler models, with slight extension: as pointed out for the compressible neutral-fluid case, and in accordance with the discussions there, we remark that the choice of the `gauge' can be more general. For instance, most obviously, $G_s$ does not need to be closed to have the same results, as long as $dG_s \wedge \Omega_s = 0$.  More essential extension will be offered in the next section.

Note that for Tao's \cite{TaoAnnPDE16} model in Eulidean space, like the above two-fluid plasma model, the two flows have their own Lagrangian maps $x=\varphi_t$ and $y=\psi_t$; that is, besides $\dot{\varphi}_t=v$, there is also $\dot{\psi}_t=u$. Ref. \onlinecite{TaoAnnPDE16} has also put down
\begin{equation}\label{eq:TaoV}
  (\partial_t + L_u)V= dp,
\end{equation}
which is a result of the no-cohomology ($b_1=0$) assumption and the Hodge decomposition for the 1-form $(\partial_t + L_u)V$ which is closed: $d[(\partial_t + L_u)V] = 0 \Leftarrow (\partial_t + L_u)\Omega = 0$ with $\Omega=dV$ and the commutation of the two operators $d$ and $\partial_t + L_u$.
Besse and Frisch \cite{BesseFrischJFM17} have written down the Cauchy invariants equation $d\dot{x}_k\wedge dy^k = \Omega_0$ corresponding to the local invariance of $\Omega$, distinguishing the two Lagrangian coordinates. We can further similarly introduce a Weber transformation function and/or a gauge to renormalize $V$ in such a way that a helicity-like 3-form $W$ is Lie invariant. Further assuming $b_3=0$ (in accordance with the Poincar\'e Lemma, say) and that $W=dT$ for some 2-form $T$, we have
\begin{equation}\label{eq:helicityTaoCauchy}
\frac{1}{2} \delta_{ij}^{kl} d T_{kl} \wedge dy^{i} \wedge dy^{j} = W_{0}.
\end{equation}
As can be seen from Eq. (\ref{eq:TaoU}), unlike the plasma two-fluid model, Tao's two-fluid model does not present a clear Lie structure, with respect to either $u$ or $v$, of (higher-order) forms for the other $u$-flow. As we will come back, the only remark is that Tao-model local structure is indeed quite different to the original Euler, though seemingly similar.

\section{`Multiplicative' local invariants from Lie-varying forms
}\label{sec:LieVarying}
We have actually applied the fact that, if the (Lie-)sources/sinks $S_{\bullet}$ are such that
\begin{equation}\label{eq:balance}
S_1 \wedge \omega_2 + \omega_1 \wedge S_2 = 0,
\end{equation}
a `multiplicative' Lie invariant follows from
\begin{equation}\label{eq:multiplicativeLie}
(\partial_t + L_v) (\omega_1 \wedge \omega_2) = S_1 \wedge \omega_2 + \omega_1 \wedge S_2 \Longleftarrow (\partial_t + L_v) \omega_i = S_i: \ i = 1, 2.
\end{equation}
%results from the Lie-varying forms
%\begin{equation}\label{eq:LieVarying}
%(\partial_t + L_v) \omega_i = S_i: \ i = 1, 2.
%\end{equation}
This is a more general and useful result than the familiar case with $S_1 = S_2= 0$, because it tells how to (Lie-)pump/damp two objects to obtain local `multiplicative' invariants. Non-trivial value also lies in the fact that some precise relations, such as the (generalised) Cauchy invariants equations, hold for local invariants, regardless the origin of the latter (from `idealness' of the flow or from some specific balance).

%\subsection{Non-ideal flow and `multiplicative' Lie invariant from Lie-varying forms}
%As noted earlier, for the non-barotropic case with $dp_s/\rho_s$ being not closed, it is generally not possible to transform $P_s$ into $\Pi_s$ such that both $\Pi_s$ and $d\Pi_s$ are invariant, thus neither the possibility of invariant local helicity from known invariant momentum and vorticity.
One important relevance of local helicity is the characterization of the degree of non-integrability of Pfaff equation (c.f., Fig. 4 of Tur and Yanovsky \cite{TurYanovskyJFM93} for a geometrical explanation) defined by the vanishing of the $1$-form and defining the surface orthogonal to the corresponding vector. Thus, global integrability, the uniform Frobenius condition, indicates null (global) helicity and results in another (higher-order) global helicity-like Godbillon-Vey invariant \cite{Hurder02}. And, we should remark that this procedure can go on and on. It is thus of our interest to control the local helicity in more general situations (of any vector in principle, but here $P_s$.)

As an example to illustrate the theoretical consideration, %of Sec. \ref{sec:LieVarying}
we now fix in the non-ideal flow the local helicity, viz., making it Lie invariant.
For simplicity of algebra and illustration, we re-write non-barotropic non-ideal two-fluid model by modeling the non-ideal effects with an exact form $dX_s$ in
\begin{equation}\label{eq:2fluidNonIdeal}
  (\partial_t + L_{v_s}) P_s = -dp_s/\rho_s + d[ (v_s,v_s)_g/2 - K_s ] + dX_s.
\end{equation}
Then, the local self-helicities, whose Hodge dual $\star(P_s \wedge dP_s)$ is also the spatial density of the global generalised helicities $\mathscr{H}_s = \int_{\mathscr{D}} \star(P_s \wedge dP_s) d\verb"vol"$, satisfy
\begin{equation}\label{eq:LocalHelicity}
  (\partial_t + L_{v_s}) P_s \wedge dP_s = [ -dp_s/\rho_s + d(v_s,v_s)_g/2 - dK_s + dX_s] \wedge dP_s - P_s \wedge d(dp_s/\rho_s).
\end{equation}
Canceling the above right-hand side, we can in principle find $dX_s$, if exists (but in general non-unique); if not, we should use a more general (not-even-closed) form to model the non-ideal effects, whose exterior derivative would appear in the Lie source/sink of $dP_s$ to complicate the calculation (simplification by working with the Hodge dual equation is possible but is not necessary for us to get into the details here), which is partly the reason why we have not let $dK_s$ absorb $dX_s$ in the above. Generalised global self-helicities have been found to be important for the dynamics of plasmas as preliminary statistical calculations for incompressible flows indicates \cite{hydrochirality}; while, the current analysis may be useful for controlling the statistics and exposing compressible fine-scale structures of the interactions of vortexes and kinetic Alfv\'en waves \cite{WuBook12}.

%Use may also be made of the above result to provide alternative model with special physical considerations. For instance, an Eulerian framework idea, arising from the nonequilibrium molecular dynamics (see, \cite{ChernovEyinkLebowitzSinaiJMP93} and references therein), is Gauss' prescription to model $S_{\bullet}$ in such a way that the system is deterministic and time reversible, with some observable, say, the kinetic energy, fixed but with the Liouville theorem invalid. Such a method, have been applied to fluids with the understanding that the dynamics with peculiar invariant(s) may be viewed as some balance in the composites of $S_{\bullet}$, ideally mode-by-mode in detail \cite{ZhuJFM16}, globally \cite{GallavottiCohenPRL95}, or intermediately for each subsystems (of selective modes, e.g., shell-by-shell subsets of Fourier modes \cite{SheJacksonPRL93}). And, now, we may have the possibility of opting to model the (Lie-)source/sink with some invariant(s) in the sense of Lie, i.e., the generalised Lagrangian framework.

An alternative angle of view to the Lie-varying form is to imagine or construct \textit{virtual} velocity fields/trajectories $\mathscr{V}$ along which the form is Lie invariant: The existence and uniqueness, besides other physical issues, of the virtual velocities for all kinds of forms deserves another study, but, as the simplest example of the pumped/damped (by the source/sink $S_{\theta}$) scalar or 0-form $\theta$, with $L_v=v \cdot \nabla$ and $L_{\mathscr{V}} = \mathscr{V} \cdot \nabla$, obviously $\mathscr{V}$ can be found with $v \cdot \nabla \theta + S_{\theta} = \mathscr{V} \cdot \nabla \theta$, except for some special location(s), if any, where $\nabla \theta = 0$ but $S_{\theta} \ne 0$. Such virtual velocities are non-unique, allowing for more (Lie-)constraints. The corresponding Lie derivatives are more complicated for other higher-order forms, and the systematic discussions are beyond the scope of this note. We remark that the problem of `multiplicative local invariant from Lie-varying forms' now becomes `multiplicative local invariants from known invariants but with the Lie derivatives being with respect to different velocities'. For example, it is obvious that in Tao's models \cite{TaoAnnPDE16}, in general $(\partial_t + L_v) \Omega \ne 0$; that is, the vorticity is not carried by its own (Biot-Savart) flow, and the other flow $u$ is such introduced that the Lie-source/sink vanishes along it. On the other hand, what Ref. \onlinecite{TaoAnnPDE16} has shown is that even though the local invariance is such preserved and both global helicity and energy conservation laws are also formally satisfied, more structures of the real Euler would be necessary to establish regularity global in time, if indeed. Thus, the accuracy of controlling the local invariance properties are important for modeling the dynamics. Here, we present the example of `fixing' (in the sense of Lie invariance) local helicity along the \textit{real} trajectories for the example of reducing Kuz'min's \cite{KuzminPLA83} corresponding (incompressible) result in flat Euclidean space to the 2D3C situation with $\partial_z = 0$, i.e., without variation in the `vertical' direction (denoted by the unit vector $\hat{z}$). 2D3C vortical modes are marked in rapidly rotating flows ubiquitous in the atmospheres of astrophysical objects, thus this example may be of strong realistic relevance. It is direct to check that now both $\theta$ (in the vertical velocity $u_z = \theta \hat{z}$) and $\zeta$ are Lie-advected by the `horizontal' velocity $u_h$ (whose curl/vorticity is $\zeta \hat{z}$), and of course by $v$. So,
\begin{eqnarray}\label{eq:LieTheta}
  (\partial_t + L_{u_h})\Theta &=& 0,
\end{eqnarray}
with $\Theta = 2\theta \zeta$.
The global helicity is also checked to read
\begin{equation}\label{eq:2D3Chelicity}
  \mathscr{H} = \int_{\mathscr{D}} \Theta d\verb"vol",
\end{equation}
invariant with such appropriate (say, periodic) boundary conditions that no boundary term appears with integration by parts. We can further check that the spatial density of $\mathscr{H}$, $\nabla \times v \cdot v$, is not $\Theta$, but the latter works in Eq. (\ref{eq:2D3Chelicity}) as if it were; so, we may use $\Theta$ as the `surrogate' of the local helicity (noting that the 0-form and 3-form local helicities are Hodge duals). Adding source/sink $S_{\bullet}$ to Lie-pump/damp both 0-forms $\theta$ and $\zeta$, respectively passive and active scalars, we have
\begin{eqnarray}
%  (\partial_t + L_{u})\theta &=& S_{\theta}, \\
  (\partial_t + L_{u_h}) f &=& S_{f}: \ f = \theta, \ \zeta.
\end{eqnarray}
Thus, setting $\theta S_{\zeta} + \zeta S_{\theta} = 0$ according to Eq. (\ref{eq:balance}), we have Eq. (\ref{eq:LieTheta}),  %$(\partial_t + L_{u_h}) \Theta = 0$ with $\Theta = \theta \zeta$,
which also provides a `Lagrangian' prescription to control the global helicity by a locally invariant $\Theta$ for such a special reduced case. In other words, we can `pretend' to take $\Theta$ to be the locally invariant density of the globally invariant helicity for some situations of studying the physical effects of controlled helicity in 2D3C or 2D passive scalar problems.

\section{Concluding remarks}
Fundamental fluid dynamics results of ideal flows can be practically useful with appropriate extensions to allow source/sink. In particular,
we have shown that the (generalised) Ohm's-law approximation, absent in our working two-fluid model, is not fundamentally crucial for plasma dynamics concerning the results of Refs. \onlinecite{BesseFrischJFM17}. The generalised Cauchy invariants equation and the Hodge-dual formula corresponding to the local helicity can arise from non-ideal flows with appropriate Lie source/sink, which may be used in constructing a physical constraint for modeling the nonequilibrium and irreversible phenomena. This latter aspect may possibly be used to compare with or go beyond the Gauss method \cite{ChernovEyinkLebowitzSinaiJMP93,GallavottiBook14} which %, though has been developed \cite{Zhu17unpublished} to be able to respect multiple constraints, in particular the conventional global helicity \cite{MoffattJFM69} with the idea of chirality-dependent dynamical eddy viscosities \cite{ZhuJFM16},
so far appears not completely clear for the general non-barotropic (plasma) flows, especially on moving domain \cite{AncoDarTufailPRSA15} for non-ideal equations in curved Riemannian manifold: Non-barotropic invariant helicities are nonlocal in time, involving the history of the Clebsch-type variable in the variational formulation, while the meaningful simple form of local invariant is possible (see, e.g., Refs. \onlinecite{WebbAncoJPA16,YahalomArxiv17} and references therein). Nevertheless, we have shown in a particular 2D3C or 2D passive scalar problem that a Lagrangian prescription is possible to control the global helicity invariance. Note that one may introduce extra freedom(s)/parameter(s) in the model of the non-ideal sink and/or source term(s) to have more than one local invariants, say, in a way similar to that pointed out in Ref. \onlinecite{ZhuJFM16} for (global) energy-helicity constraints of turbulence modeling. The local constraint may be useful in the sense that when a small-scale plasma becomes so turbulent that only measure-valued solutions could be considered, modeling such `micro-turbulence' with local constraints from the properties of the ideal classical solution will make whatever approximation less uncontrollable. Such considerations are also supported by our further analysis in a two-fluid fashion of the local structures of Tao's \cite{TaoAnnPDE16} modification of the Euler equation which keeps some seemingly same but fundamentally quite different local invariance laws.

Although only some formal examples are offered to outline the calculations, physical relevance with realistic flows is particularly interestingly reflected in the two-dimensional-three-component (2D3C) or the 2D passive scalar problem. The locally invariant $\Theta = 2\theta \zeta$, with $\theta$ and $\zeta$ being respectively the scalar value of the `vertical velocity' (or the passive scalar) and the `vertical vorticity', may be used as if it were the spatial density of the globally invariant helicity, providing a Lagrangian prescription to control the latter in some situations of studying its physical effects in rapidly rotating flows (ubiquitous in atmosphere of astrophysical objects) with marked 2D3C vortical modes or in purely 2D passive scalars.

In the end, we would like to remark that, although we can gauge $\omega_i$ in Eq. (\ref{eq:multiplicativeLie}) to some Lie invariant form, it is a completely different issue, because the purpose is really `constructing' some `peculiar' invariant of particular modeling interest or of physical importance, rather than `passively' finding new ones.

\begin{acknowledgements}
\begin{CJK*}{GB}{gbsn}
%\begin{CJK*}{KS}{}
%\begin{CJK*}{GBK}{kai}

This work is supported by NSFC (No. 11672102) and ��԰ѧ�� fundings. The help from A. Elbakyan is also acknowledged.

%\end{CJK*}
%\end{CJK*}
\end{CJK*}
\end{acknowledgements}

\end{document}